\documentclass[showpacs,amsmath,nofootinbib,amssymb, reprint,prd]{revtex4-1}
\pdfoutput=1
\usepackage{mathtools}
\mathtoolsset{showonlyrefs}
\usepackage{psfrag}
\usepackage{array}
\usepackage{amssymb}
\usepackage{amsmath}
\usepackage{amsthm}
\usepackage{graphicx}
\usepackage{caption}
\usepackage[labelsep=quad]{subcaption}
\usepackage{epstopdf}
	
\usepackage{epsfig}
\usepackage[punctsep]{collref}
\collectsep[]{;}

\def\shil{ {\cal H}_{\Psi}}

\def\al{A}

\def\pb[#1,#2]{\{#1, #2\}}
\def\deb[#1,#2]{[#1,#2]_{\text{D.B.}}}

\def\tO{\widetilde{\cal O}}

\def\tr{{\rm Tr}}

\def\tb{\widetilde{b}}

\def\Or[#1]{{\text{O}}\left({#1}\right)}
\def\dotl[#1,#2]{\left\langle #1,\, #2 \right\rangle}
\def\dotlb[#1,#2]{\left\langle #1,\, #2 \right\rangle}
\def\dotlm[#1,#2]{\left[ #1,\, #2 \right]}
\def\dotp[#1,#2]{(\vect{#1} \cdot\vect{#2})}
\def\aff[#1,#2]{\hat{#1}(#2)}
\def\nc{{\cal N}}
\def\n4sym{{\cal N}=4 SYM}
\def\>{\rangle}
\def\<{\langle}
\def\weight[#1,#2,#3]{\{(#1),#2,#3\}}
\def\ads[#1]{$\text{AdS}_{#1}$}

\newcommand{\be}{\begin{equation}}
\newcommand{\ee}{\end{equation}}
\newcommand{\ba}{\begin{align}}
\newcommand{\ea}{\end{align}}
\newcommand{\bs}{\begin{split}}
\def\sess\end{split}
\newcommand{\vect}[1]{{\boldsymbol{#1}}}

\begin{document}
\title{The Black Hole Interior in AdS/CFT and the Information Paradox}
\author{Kyriakos Papadodimas}
\email{kyriakos.papadodimas@cern.ch}
\affiliation{Centre for Theoretical Physics, University of Groningen, Nijenborgh 4, 9747 AG, The Netherlands }
\affiliation{Theory Group, Physics Department, CERN, CH-1211 Geneva 23,
Switzerland.}
\author{Suvrat Raju}
\email{suvrat@icts.res.in}
\affiliation{International Centre for Theoretical Sciences, Tata Institute of Fundamental Research, IISc Campus, Bengaluru 560012, India.}
\begin{abstract}
We show that, within the AdS/CFT correspondence, recent formulations of the information paradox can be reduced to a question about the existence of certain kinds of operators in the CFT. We describe a remarkably simple construction of these operators on a given state of the CFT. Our construction leads to a smooth
horizon, addresses the strong subadditivity paradox, while preserving locality within
effective field theory, and reconciles the existence of the interior with the growth of states with energy in the CFT. We also extend our construction to non-equilibrium states.   
\end{abstract}
\maketitle
\section{Introduction}
The past few months have seen an intense renewal of interest in the information paradox \cite{Almheiri:2012rt,Almheiri:2013hfa, Marolf:2013dba,Mathur:2009hf}. Although the AdS/CFT correspondence \cite{Maldacena:1997re,Witten:1998qj,Gubser:1998bc} strongly suggests that
the formation and evaporation of a black hole should be unitary,
several
authors have recently argued that this is inconsistent with a
smooth horizon and interior for the black hole. These arguments
can essentially be summarized as the claim that the CFT does not contain operators that can describe the interior.  In this paper, we 
present an explicit and strikingly simple construction for such operators, which ensures that they have the right
properties on a given state of the CFT and on its descendants produced by acting
on this state with other light operators.

 This construction successfully addresses all the recent arguments in favour of structure at the black hole horizon. So, the central thrust of this paper is to show that if one is allowed to use different operators to describe the interior of the black hole in different states of the CFT---this naturally aligns with the expectation that one cannot define ``background-independent'' local observables in a theory of quantum gravity---then there is no need for firewalls at the horizon.

Before proceeding to our construction, we briefly describe the observables that we are interested in. In the appropriate regime of parameters, the bulk
theory of quantum gravity is approximately local. This means that the boundary CFT should have operators $\phi^i_{\text{CFT}}(\vect{x})$, where $\vect{x}$ can be interpreted as a bulk point, with the property that the CFT correlation
functions $\langle \Psi | \phi^{i_1}_{\text{CFT}}(\vect{x_1}) \ldots \phi^{i_n}_{\text{CFT}}(\vect{x_n}) |\Psi \rangle$ should be the same as those obtained 
from effective field theory on the geometry dual to the state $|\Psi \rangle$. 

However, these ``local'' observables have an important limitation. If we take two points very close $|\vect{x}_i - \vect{x}_j| < l_{\text{pl}}$ or take
the number of points $n$ to scale with the central charge\footnote{Note that ${\cal N} \propto P^2$ in the supersymmetric $SU(P)$ theory} of the CFT,  ${\cal N}$, then the correlators cease to have a semi-classical bulk interpretation. This is a generic statement about local observables in quantum gravity, and the reader
should always keep this in mind. 

It is well known that one can identify operators in the CFT that describe
the exterior of a bulk black hole, as we review in \S \ref{sec:localops}.  In this paper, we find operators, which depend on the state $|\Psi \rangle$, that describe the interior as well.  After describing this construction for equilibrium states in \S \ref{sec:constructduals}, we use it to address recent arguments for structure at the horizon in \S \ref{sec:resolveinfo}.  
We describe how to to extend our construction to non-equilibrium states in \S \ref{sec:noneq}.

\section{Bulk Local Operators from the CFT \label{sec:localops}}
We consider a CFT on $S^{d-1} \times R$, with a large central charge ${\cal N}$, placed in a pure state $|\Psi \rangle$ that is expected to be dual
to an AdS-Schwarzschild black hole in the bulk. The CFT contains several light
local operators  ${\cal O}^i(t, \Omega)$, distinguished by the index $i$. The AdS/CFT dictionary tells us that the ``single trace'' operators in this set are dual
to fields in the bulk. 

We start by considering the case where the state $|\Psi \rangle$ is in equilibrium. This implies that correlators in this state can be approximated, at large $\nc$, by thermal correlators
\be
\label{thermalstate}
\begin{split}
&\langle \Psi | {\cal O}^{i_1}(t_1, \Omega_1) \ldots {\cal O}^{i_n}(t_n, \Omega_n) | \Psi \rangle \\ &= Z_{\beta}^{-1}\tr\big(e^{-\beta H} {\cal O}^{i_1}(t_1, \Omega_1) \ldots {\cal O}^{i_n}(t_n, \Omega_n)\big).
\end{split}
\ee
$Z_{\beta}$ is the partition function of the CFT at temperature $\beta^{-1}$ associated with $|\Psi \rangle$, and $H$ is the CFT Hamiltonian. If the
state is charged under a conserved charge $\hat{Q}$, then we can associate a chemical potential $\mu$ with it, and should replace $\beta H \rightarrow \beta H + \mu \hat{Q}$ everywhere below.

Under these conditions, it is possible to write down an explicit operator in the CFT that can be interpreted as this bulk local field. We showed in \cite{Papadodimas:2012aq} , following
\cite{Banks:1998dd,  *Bena:1999jv, *Hamilton:2006az, *Hamilton:2005ju, *Hamilton:2007wj, *VanRaamsdonk:2009ar, *VanRaamsdonk:2010pw,*VanRaamsdonk:2011zz}, that this operator
could be written as
\be
\label{cftthermal}
\phi_{\text{CFT}}^i (t,\Omega, z) = \sum_{m,n}   
\, \big[ {\cal O}^i_{\omega_n, \vect{m}} f_{\omega_n, \vect{m}}(t, \Omega, z) + \text{h.c.} \big],
\ee
outside the black hole.
Here ${\cal O}^i_{\omega_n, \vect{m}}$ are the modes of the local operator
on the sphere, and in time, and $f$ is an appropriately chosen function. In the CFT, $(t,\Omega,z)$ are just labels for the operator, but they have an
interpretation as coordinates in \ads[d+1].
This formula can be corrected order by order in ${1 \over \nc}$ \cite{Kabat:2012av}.

We have discretized the frequencies $\omega_n$. This can be done by placing the boundary theory on a lattice, and then considering a discrete Fourier transform, and another method is discussed in \cite{Papadodimas:2013}. Although this discretization is necessary, its precise details are not relevant, and the the reader may choose to think of the boundary theory on a lattice with $e^{\sqrt{\nc}} = e^P$ points in the $SU(P)$ theory, which certainly gives as close an approximation to the continuum theory as one might desire.

Repeating this analysis for the black hole interior we find that we require new operators $\tO^{i}$, 
\be
\label{interiorexpansion}
\begin{split}
\phi_{\text{CFT}}^i(t,\Omega,z) =
&\sum_{\vect{m},n}  \big[ {\cal O}^i_{\omega_n,\vect{m}}\,g_{\omega_n,\vect{m}}^{(1)}(t,\Omega,z) \\ &+ \widetilde{\cal O}^i_{\omega_n,\vect{m}} \,g_{\omega_n,\vect{m}}^{(2)}(t,\Omega,z)+ \text{h.c.} \big],
\end{split}
 \ee
where the functions $g^{(1)}$ and $g^{(2)}$ can be written down explicitly \cite{Papadodimas:2012aq}. The ``mirror operators'' $\widetilde{\cal O}^i$ must have the following important property
\begin{widetext}
\be
\label{otildewithincorr}
\begin{split}
 &\langle \Psi| {\cal O}^{i_1}(t_1,\Omega_1)\ldots\widetilde{\cal O}^{j_1} (t_1',\Omega_1')\ldots \tO^{j_l} (t_{l}', \Omega'_{l})\ldots  {\cal O}^{i_n} (t_{n}, \Omega_{n}) |\Psi \rangle \\ &= Z_{\beta}^{-1}{\rm Tr}\left[e^{-\beta H} {\cal O}^{i_1} (t_1,\Omega_1)\ldots{\cal O}^{i_n}(t_n, \Omega_n) {\cal O}^{j_{l}} (t_{l}' + i \beta/2, \Omega_{l}') \ldots {\cal O}^{j_1}\left(t_1' + i\beta /2,\Omega_1'\right)  \right],
\end{split}
\ee
\end{widetext}
where, as shown,  all the analytically continued operators are moved to the right of the ordinary operators in the trace, and their ordering is reversed. This equation holds at large $\nc$.

If operators $\widetilde{\cal O}^i(t,\Omega)$ that obey \eqref{otildewithincorr} can be found in the CFT then one can explicitly compute a correlation function across the horizon of a black hole and show that it is smooth \cite{Papadodimas:2012aq}, since
this correlator reduces to a calculation in the eternal black hole 
of \cite{Maldacena:2001kr}. So, the 
question of whether the black hole interior can be described within AdS/CFT is entirely one of finding operators that obey \eqref{otildewithincorr} in the CFT. 

\section{Mirror Operators in an Equilibrium State \label{sec:constructduals}}
We now find such operators, and use them to address all the recent arguments on the information paradox.

Consider the set of polynomials in the modes of 
CFT operators
\be
\al_{\alpha} = \sum_{N}  \alpha(N) ({\cal O}^i_{\omega_n, \vect{m}})^{N(i,n,\vect{m})},
\ee
where $\alpha(N)$ are arbitrary coefficients, and the sum runs over all functions $N$, with the important restrictions that 
\be
\label{energybound}
\begin{split}
&\sum_{i,n,\vect{m}} N(i,n,\vect{m}) \omega_n  \leq E_{\text{max}} \ll \nc.
\end{split}
\ee
This set of polynomials forms a linear space, and we also bound the number of insertions, $\sum N(i,n,m) \leq K_{\text{max}}$, to limit the dimension
of this space: ${\cal D}_{\cal A} \ll e^{\nc}$. 

We consider this set modulo all operator relations in the CFT. For example, if we have two operators with $[{\cal O}_1, {\cal O}_2] = 1$ then, of course, the polynomial ${\cal O}_1 {\cal O}_2$ is identified with ${\cal O}_2 {\cal O}_1 + 1$.
These polynomials can also be thought of as suitably regularized polynomials of local operators in the CFT. We exclude the zero-modes of conserved currents from these polynomials and return to them later.

Recalling the restrictions on the possible observables in quantum gravity that we mentioned above,
the most general set of observables, for which we can expect a semi-classical bulk interpretation are the expectation values $\langle \Psi | \al_{\alpha} | \Psi \rangle$. 

It is important that a generic state $|\Psi \rangle$ with
$\langle \Psi | H | \Psi \rangle = \Or[\nc]$ also satisfies
\be
\label{noannihilpoly}
\al_{\alpha} |\Psi \rangle \neq 0, \qquad  \forall A_\alpha \neq 0.
\ee
This follows simply because ${\cal D}_{\cal A} \ll e^{\nc}$, and there are of order $e^{\nc}$ states at energy $\nc$. In fact, we can even take \eqref{noannihilpoly} as part of a definition of what we mean by a generic state.

Now, we can think of the vector space $\shil = \al_{\alpha} |\Psi \rangle$, formed by acting with all possible polynomials on the state $|\Psi\rangle$. Then \eqref{noannihilpoly} tells us that the set of observable polynomials and $\shil$ are {\em isomorphic} as linear vectors spaces. 
Moreover, the only relevant observables are then $\langle \Psi | v \rangle$, where $|v\rangle \in \shil$. 

We now {\em define} the mirror operators $\tO^{i}_{n,\vect{m}}$ by their action on $\shil$
\be
\label{todefpoly}
\tO^i_{\omega_n,\vect{m}} \al_{\alpha} |\Psi \rangle =  \al_{\alpha} e^{-{\beta \omega_n \over 2}} ({\cal O}^{i}_{\omega_n,\vect{m}})^{\dagger}|\Psi \rangle.
\ee
This simple
definition will turn out to have remarkable properties.

First, note that, since the polynomials form a linear space, and \eqref{todefpoly}
is also linear, we can equivalently define the $\tO^{i}_{n,\vect{m}}$ operators
by their action on a basis of $\shil$, which comprises ${\cal D}_{\cal A}$ linearly independent vectors. It is {\em always} possible to find a linear
operator with any specified action on a 
linearly independent set of vectors. So, we can find operators $\tO^i_{n,\vect{m}}$ in the CFT that satisfy \eqref{todefpoly}. 

In fact \eqref{todefpoly} does not uniquely specify $\tO^i_{n, \vect{m}}$, since
their action outside $\shil$ is unspecified but, as we will show, this ambiguity is unimportant except in very high order correlators. 

Let us give another, completely equivalent, way of defining these operators. Consider the anti-linear map from $\shil \rightarrow \shil$
\be
S \al_{\alpha} |\Psi \rangle = \al_{\alpha}^{\dagger} |\Psi \rangle.
\ee
Note that $S^2 = 1$ and $S|\Psi \rangle = |\Psi \rangle$. 
The existence of such a map follows from the reasoning above. Then the mirror operators are simply defined by 
\be
\label{todefwiths}
\tO^i_{\omega_n, \vect{m}} = S  e^{-{\beta H \over 2}} {\cal O}^{i}_{\omega_n, \vect{m}} e^{\beta H \over 2} S,
\ee
This is precisely the map that appears in the Tomita-Takesaki theory of
modular automorphisms of von Neumann algebras, as we explore in \cite{Papadodimas:2013}.

Let us verify that these operators do satisfy the crucial relation \eqref{otildewithincorr}. We find
\begin{widetext}
\be
\label{otildewithincorrmom}
\begin{split}
\langle \Psi| {\cal O}^{i_1}_{\omega_1, \vect{m}_1}  \ldots \widetilde{\cal O}^{j_1} _{\omega_1', \vect{m}_1'} \ldots \widetilde{\cal O}^{j_{l}}_{\omega_{l}', \vect{m}_{l}'} \ldots {\cal O}^{i_n}_{\omega_n, \vect{m}_n}|\Psi \rangle =  e^{-{\beta \over 2} (\omega_1' + \ldots \omega_{l}')} 
 \langle \Psi| {\cal O}^{i_1}_{\omega_1, \vect{m}_1} \ldots {\cal O}^{i_n}_{\omega_n, \vect{m}_n} ({\cal O}^{j_{l}}_{\omega_{l}', \vect{m}_{l}'})^{\dagger} \ldots ({\cal  O}^{j_1}_{\omega_1', \vect{m}_1'})^{\dagger} | \Psi \rangle.
\end{split}
\ee
\end{widetext}
Here, we use the definition \eqref{todefpoly} recursively to pass the $\tO$ operators to the right, and convert them to ordinary operators in turn. 
Alternately,
the reader may verify this relation using \eqref{todefwiths}. Fourier
transforming back to position space, taking care that $\tO_{\omega_n,\vect{m}}$ 
has energy $-\omega_n$ by \eqref{todefpoly}, and using \eqref{thermalstate},  we find that \eqref{otildewithincorr} holds!

Now we turn to two subtleties. First, in proving \eqref{otildewithincorrmom},  we tacitly assumed that   products of operators always kept us within $\shil$. If we are considering a very high point correlator, on the edge of the 
bound \eqref{energybound}, then acting with the $\tO$-operators repeatedly could take us out of the space $\shil$ on which \eqref{todefpoly} holds.  Clearly, these ``edge'' effects are important only for very high-point correlators, and we can make them small enough to be unimportant at {\em any order} in the ${1 \over {\cal N}}$ expansion, by taking $E_{\text{max}}$ large enough, but still much smaller than $\nc$.

Second, we turn to conserved charges, or the zero-modes of conserved currents. We denote a polynomial in these charges, including the Hamiltonian, by  ${\cal Q}_{\alpha}$. 
If $|\Psi \rangle$ is an energy eigenstate, or transforms in a small representation of the
symmetry group, it could be annihilated by the action of some ${\cal Q}_{\alpha}$. We generalize \eqref{noannihilpoly} to demand that acting by $\al_{\gamma_i}$  generates no further linear dependencies
\be
\sum_i \al_{\gamma_i} {\cal Q}_{\alpha_i} |\Psi \rangle = 0 \Rightarrow \exists \kappa_i \in \mathbb{C}, ~\text{s.t.}~ \sum_i \kappa_i {\cal Q}_{\alpha_i} |\Psi \rangle = 0.
\ee
As above, we can write any state as a linear combination of states where  ${\cal Q}_{\alpha}$ acts immediately on $|\Psi \rangle$, by commuting it through the $\al_{\gamma}$. We now define the mirror 
through 
\be
\label{cftgaugeinvar}
\tO^i_{\omega_n, \vect{m}}
\al_{\gamma} {\cal Q}_{\alpha} |\Psi \rangle = \al_{\gamma} e^{-{\beta \omega_n \over 2}} ({\cal O}^{i}_{\omega_n,\vect{m}})^{\dagger} {\cal Q}_{\alpha} |\Psi \rangle.
\ee
This ensures that if an operator 
transforms in some representation, then  its mirror transforms in the conjugate representation.
The action of $S$ can also be suitably generalized \cite{Papadodimas:2013}.

\section{Obtaining a smooth horizon \label{sec:resolveinfo}}
Now, we show how the use of these operators resolves all recent arguments that suggest the presence of firewalls or fuzzballs at the horizon of a black hole. 

We start with the strong subadditivity paradox \cite{Mathur:2009hf, Almheiri:2012rt}. Consider three regions of an evaporating black hole: $B$, just outside the horizon, $E$, which is far away, and $\widetilde{B}$, just behind the horizon. The horizon is smooth, only if the modes in $B$ are entangled
with the modes in $\widetilde{B}$; this is also expressed by \eqref{otildewithincorr}. On the other hand, after the Page time \cite{Page:1993df} of the black hole, the modes in $B$ must also be entangled with the modes in $E$. 

We phrase this precisely in CFT language in \cite{Papadodimas:2013}, but
the strong subadditivity of entropy now tells us that the modes in $E$ cannot be independent of the modes in $\widetilde{B}$. 
In terms of local operators, this means that for any function $f(\vect{x})$ localized on a nice slice in region $\widetilde{B}$ , we can find a function $g(\vect{x})$, localized on the {\em same} nice slice in region $E$, so that
\be
C = [\int \phi_{\text{CFT}}^i(\vect{x}) f(\vect{x}) d^{d} x , \int \phi_{\text{CFT}}^i(\vect{y}) g(\vect{y}) d^d y] \neq 0.
\ee

Black hole complementarity  \cite{'tHooft:1990fr, *Susskind:1993if} leads us to expect that the degrees of freedom inside the black hole are {\em not} independent of the degrees of freedom outside but the trick is to simultaneously preserve effective locality, which seems to be violated by the non-zero commutator above.

However, in our construction, this lack of independence is cleverly hidden
so that $C$, or even its powers like $C^{\dagger} C$ can {\em never} be detected in low-point correlators. In particular we have
\be
\label{commutvanish}
\langle \Psi | \al_{\alpha_1} C^{\dagger} C \al_{\alpha_2} |\Psi \rangle = 0,
\ee
as long as the product polynomial $\al_{\alpha_1} \al_{\alpha_2}$ also satisfies the restriction \eqref{energybound}. This follows because, using \eqref{interiorexpansion}, we can translate \eqref{commutvanish} into a statement about the commutators of $\tO^i$ with ${\cal O}^i$. Even though this commutator 
does not vanish as an operator, we proved above that it is undetectable within low-point correlators. This is obvious from substituting the commutator in \eqref{otildewithincorrmom}, and also follows directly from \eqref{todefpoly}. This commutator can be detected in a correlator where the 
number of points scales with $\nc$, but as advertised, we should not expect semi-classical bulk-locality to hold in any sense for such correlators.

Next, consider the ``lack of a left-inverse'' paradox \cite{Almheiri:2013hfa}. With $G =  \langle \Psi | [{\cal O}^i_{\omega_n, \vect{m}}, ({\cal O}^i_{{\omega_n}, \vect{m}})^{\dagger}] |\Psi \rangle$, define the shorthand operators
\be
\begin{split}
&b = G^{-{1 \over 2}} {\cal O}^i_{\omega_n, \vect{m}}; \quad b^{\dagger} = G^{-{1 \over 2}} ({\cal O}^i_{\omega_n, \vect{m}})^{\dagger}, \\
&\widetilde{b} = G^{-{1 \over 2}} \tO^i_{\omega_n,\vect{m}}; \quad \widetilde{b}^{\dagger} = G^{-{1 \over 2}} (\tO^i_{\omega_n,\vect{m}})^{\dagger}.
\end{split}
\ee
We note that
\be
\label{commoneonstate}
[\widetilde{b}, \widetilde{b}^{\dagger}] \al_{\alpha} |\Psi \rangle = \al_{\alpha} [b, b^{\dagger}] |\Psi \rangle = \al_{\alpha} |\Psi \rangle + \Or[\nc^{-1}].
\ee
However, since $[H, \widetilde{b}^{\dagger}] = -\omega_n b^{\dagger}$, we cannot have $[\widetilde{b}, \widetilde{b}^{\dagger}] = 1$ as an operator equation. This would have required $\widetilde{b}^{\dagger}$ to have a left-inverse, which it cannot since it maps states of higher energy to lower energy states, and the number of states grows monotonously with energy. 

In our case, it is clear that $[\widetilde{b}, \widetilde{b}^{\dagger}] \neq 1$, since \eqref{commoneonstate} holds only when $\al_{\alpha}$ satisfies
\eqref{energybound}. So $\widetilde{b}^{\dagger}$ can have null vectors. These operators have a commutator that is effectively $1$
when inserted in any low-point correlator but differs from $1$ as an operator. This difference is only detectable in a correlator where the energy
of insertions scales with $\nc$.
 
Finally, consider the number operator measured by the infalling observer.
\be
\begin{split}
N_a = (1 -e^{- \beta \omega_n})&^{-1} \Big[ \left(b^{\dagger} - e^{-{\beta \omega_n\over 2} }\tb \right) \left(b - e^{-{\beta \omega_n\over 2} } \tb^{\dagger} \right) \\ &+ \left( \tb^{\dagger} - e^{-{\beta \omega_n\over 2} } b \right) \left( \tb - e^{-{\beta \omega_n\over 2} }b^{\dagger} \right)  \Big].
\end{split}
\ee
We see immediately by virtue of \eqref{todefpoly} that $N_a |\Psi \rangle = 0$. The argument of \cite{Marolf:2013dba} that $N_a \neq 0$ in a typical state can basically be summarized as follows. For generic operators
$b, b^{\dagger}, \tb, \tb^{\dagger}$ we do not expect $\big(\tb - e^{-{\beta \omega_n\over 2} }b^{\dagger} \big) |\Psi \rangle = \left(b - e^{-{\beta \omega_n\over 2} } \tb^{\dagger} \right) |\Psi \rangle = 0$ to hold, since there is no ``energetic'' reason for it. However, our operators
are precisely selected in a state-dependent manner to satisfy this relation, and this undercuts the argument.

In a different language, this argument that state-independent operators cannot be ``entangled'' in a generic state also underpins the ``theorem'' \cite{Mathur:2009hf} that
small-corrections cannot unitarize radiation. Our state-dependent construction allows
correlators outside and across the horizon to be very close to their effective field theory expectations,
within the unitary
framework of the CFT.

\section{Non-Equilibrium Scenarios \label{sec:noneq}}
We can already study time-dependent correlation functions on an equilibrium state. This includes problems where the equilibrium
black hole background is excited by some sources. In such a setting, since $N_a \al_{\alpha} |\Psi \rangle \neq 0$, we see that the infalling observer would notice particles. However, now we turn to a setting where the base state  on which we define the mirror operators is out of equilibrium.

An equilibrium state is defined \cite{Papadodimas:2013} as a state where if we consider $\chi_{\alpha}(t)  = \langle \Psi | e^{i H t} \al_{\alpha} e^{-i H t} |\Psi \rangle$  then ${1 \over T} \int_0^T |\chi_{\alpha}(t) - \chi_{\alpha}(0)|$ is exponentially small in $\nc$,  where the time $T$ is taken to be the inverse of the smallest difference in our discrete frequencies: $T = (\omega_{n+1} -\omega_{n})^{-1}$.

We consider a class of ``near-equilibrium'' states produced by 
exciting an equilibrium state by turning on a source for some CFT operators. More precisely, where $\al_{\alpha}$ is a Hermitian polynomial, we consider states
\be
\label{outofeq}
|\Psi' \rangle = U |\Psi \rangle, \quad U = e^{i \al_{\alpha}},
\ee

Clearly, if we wanted to define the mirror operators by using $|\Psi' \rangle$ as a base-state, we cannot just use \eqref{todefpoly}. Instead, 
we ``strip'' off the excitation $U$, and  use the mirror operators
on the equilibrium state $|\Psi \rangle$. 

In fact, we can show that given a near-equilibrium state $|\Psi'\rangle$, it can be written uniquely in the form \eqref{outofeq}: the unitary $U$ such that $U^{\dagger} |\Psi' \rangle$ is in equilibrium is fixed by the state $|\Psi' \rangle$ itself. 
Intuitively this is clear. Once we have found a unitary that takes $|\Psi' \rangle$ to an equilibrium state, turning on any other source would only
take it out of equilibrium again.

So, given a near-equilibrium state $|\Psi' \rangle$, we define
the mirror operators by
\be
\widetilde{\cal O}^i_{\omega_n,\vect{m}} \al_{\alpha} |\Psi' \rangle = \al_{\alpha} U e^{-{\beta \omega_n \over 2}} ({\cal O}^{i}_{\omega_n, \vect{m}})^{\dagger} U^{\dagger} |\Psi' \rangle. 
\ee
Equivalently, we can write 
\be
\tO^i_{\omega_n,\vect{m}} = S U e^{\beta H \over 2} {\cal O}^i_{\omega_n, \vect{m}} e^{-{\beta H \over 2}} U^{\dagger} S,
\ee
for near-equilibrium states.

It is clear that this prescription simply converts correlators on $|\Psi' \rangle$ into correlators on $|\Psi \rangle$ with an additional excitation $U$. Since these correlators correctly reproduce all kinds of correlators on a black hole in equilibrium, the correlators on $|\Psi'\rangle$ meet the expectations for semi-classical correlators in an excited state of the black hole. This successfully addresses the ``frozen vacuum'' argument \cite{Bousso:2013ifa}.

\section{Conclusions and Discussion}
In this paper, we have argued that AdS/CFT can describe the interior of the black hole and predicts a smooth horizon, precisely in line with semi-classical expectations. The principle underlying this paper is that one should expect a good semi-classical interpretation only for correlation functions, where the total  energy of the operator insertions does not scale with $\nc$. A super-observer in the CFT may measure more complicated correlators, but these do not necessarily have a local bulk description. 

We constructed mirror operators that depended on the state of the theory, and behaved correctly when evaluated on this state,
or its descendants produced by acting on it with light operators. The arguments of \cite{Almheiri:2012rt,Almheiri:2013hfa, Marolf:2013dba} suggest that such state-dependent
operators are necessary to describe the interior. This feature
also appeared in our previous work \cite{Papadodimas:2012aq} and in \cite{Verlinde:2013uja,*Verlinde:2012cy,*Verlinde:2013vja}. This is not unexpected  in light of the belief that it is impossible to localize operators in quantum gravity in a background-independent manner.  

Given the surprising power of  state-dependent operators, in resolving every one of  the recent issues surrounding
the information paradox, it is clear that this issue of state dependence
in quantum gravity is an important area for further investigation.

\section*{Acknowledgments}
We are grateful to all the participants of the CERN institute on ``Black Hole Horizons and Quantum Information'', the seventh regional meeting on string theory at Crete, the KITP ``Fuzz, Fire or Complementarity'' workshop, the ICTS discussion meeting on the ``Information Paradox, Entanglement and Black Holes'', and the ``Cosmological Frontiers in Fundamental Physics'' conference at the Perimeter Institute. We are also grateful to Rajesh Gopakumar, Shiraz Minwalla and Lubo\v{s} Motl for comments on a draft of this manuscript. K.P. acknowledges support
from the Royal Netherlands Academy of Arts and Sciences (KNAW). S.R. is partially supported by a
Ramanujan fellowship of the Department of Science and Technology (India). 

\bibliography{references}
\end{document}